\documentclass[aps,12pt,final,notitlepage,oneside,onecolumn,
nobibnotes,nofootinbib,superscriptaddress,noshowpacs,centertags]
{revtex4}

\usepackage{graphicx}
\usepackage{amssymb}

\newcommand{\beq}{\begin{equation}}
\newcommand{\eeq}{\end{equation}}
\newcommand{\bea}{\begin{eqnarray}}
\newcommand{\eea}{\end{eqnarray}}
\def\eps{\varepsilon}

\begin{document}
%\selectlanguage{english}

\title{
Isothermal vs.~isentropic description of protoneutron stars 
in the Brueckner-Bethe-Goldstone theory}

\author{G.F. Burgio and H.-J. Schulze}
\affiliation{
INFN, Sezione di Catania, Via Santa Sofia 64, I-95123 Catania, Italy}

\begin{abstract}
We study the structure of hadronic protoneutron stars 
within the finite temperature Brueckner-Bethe-Goldstone theoretical approach.
Assuming beta-equilibrated nuclear matter with nucleons and leptons in
the stellar core, with isothermal or isentropic profile, we 
show that particle populations and equation of state are very similar. 
As far as the maximum mass is concerned, we find that its value 
turns out to be almost independent on T, while a slight decrease is observed 
in the isentropic case, due to
the enhanced proton fraction in the high density range.
\end{abstract}

\maketitle

\section{Introduction}

A protoneutron star (PNS) is formed after a successful supernova explosion 
and constitutes for several tens of seconds a transitional state 
to either a neutron star or a black hole \cite{pra}.
Initially, the PNS is optically thick to neutrinos, that is, they
are temporarily trapped within the star. 
The subsequent evolution of the PNS is dominated by neutrino diffusion,
which first results in deleptonization and subsequently in cooling. 
After a much longer time, photon emission competes with neutrino
emission in neutron star cooling.

In this paper, we will focus upon the essential ingredient that
governs the macrophysical evolution of neutron stars, i.e., 
the equation of state (EOS) of dense matter at finite temperature. 
We have developed a microscopic EOS
in the framework of the Brueckner-Bethe-Goldstone (BBG) many-body approach 
extended to finite temperature. 
This EOS has been successfully applied to the study 
of the limiting temperature in nuclei \cite{big1,big2}.
The scope of this work is to present results on the composition and structure 
of these newly born stars with the EOS previously mentioned.

Two new effects have to be considered for a PNS. 
First, thermal effects which result in entropy production, with values
of a few units per baryon, and temperatures up to 30-40 MeV \cite{burr,pons}. 
Second, the fact that neutrinos are trapped in the star, which means that 
the neutrino chemical potential is non-zero.
This alters the chemical equilibrium and leads to compositional changes. 
Both effects may result in observable consequences in the neutrino signature 
from a supernova and may also play an important role in determining whether 
or not a given supernova ultimately produces a cold neutron star or a 
black hole. 

This paper is organized as follows. 
In Section 2 we briefly illustrate the BBG
many-body theory at finite temperature. 
Section 3 is devoted to the study of the 
stellar matter composition at constant temperature or entropy, 
and the resulting EOS. 
In Section 4 we discuss our results regarding the structure of 
(proto)neutron stars, in particular their maximum mass. 
Finally, conclusions are drawn in Section 5.

\section{The BBG theory at finite temperature}

In the recent years, the Brueckner-Bethe-Goldstone perturbative 
theory \cite{book}
has been extended in a fully microscopic way to the finite temperature case
\cite{big1},
according to the formalism developed by Bloch and De Dominicis \cite{bloch}. 
In this approach the
essential ingredient is the two-body scattering matrix $K$, which,
along with the single-particle potential $U$, satisfies the
self-consistent equations
\begin{eqnarray}
  \langle k_1 k_2 | K(W) | k_3 k_4 \rangle
 &=& \langle k_1 k_2 | V | k_3 k_4 \rangle
\nonumber\\&&
 \hskip -30mm +\; \mathrm{Re}\!\sum_{k_3' k_4'}
 \langle k_1 k_2 | V | k_3' k_4' \rangle\!
 { [1\!-n(k_3')] [1\!-n(k_4')] \over
   W - E_{k_3'} - E_{k_4'} + i\epsilon }
 \langle k_3' k_4' | K\!(\!W\!) | k_3 k_4 \rangle
%\nonumber\\&& 
\label{eq:kkk}
\end{eqnarray}
and
\begin{equation}
 U(k_1) = \sum_{k_2} n(k_2) \langle k_1 k_2 | K(W) | k_1 k_2 \rangle_A \:,
\label{eq:ueq}
\end{equation}
where $k_i$ generally denote momentum, spin, and isospin. Here $V$
is the two-body interaction,
$W = E_{k_1} + E_{k_2}$ represents the
starting energy, $E_k = k^2\!/2m + U(k)$ the single-particle
energy, and  $n(k)$ is the finite temperature Fermi distribution. 
We chose the Argonne $V_{18}$ nucleon-nucleon potential 
\cite{wiringa} as two-body interaction $V$, supplemented by
three-body forces (TBF) among nucleons, 
in order to reproduce correctly the nuclear
matter saturation point $\rho_0 \approx 0.17~\mathrm{fm}^{-3}$, $E/A
\approx -16$ MeV. We have adopted the phenomenological Urbana model
\cite{schi}, which consists of an
attractive term due to two-pion exchange with excitation of an
intermediate $\Delta$ resonance, and a repulsive phenomenological
central term. In the BBG approach, the TBF is reduced to a 
density-dependent two-body force by averaging over the position of the third
particle \cite{bbb}.

In order to simplify the numerical procedure, we introduce the so-called
{\em Frozen Correlations Approximation}, 
i.e., the correlations at $T\neq 0$ are assumed to be essentially 
the same as at $T=0$. 
This means that the single-particle potential $U_i(k)$ for the component $i$ 
can be approximated by the one calculated at $T=0$. 
The accuracy of that approximation has been checked in Ref.~\cite{big1}.

Within this approximation, for a fixed density  $\rho= \sum_k n(k)$ and
temperature $T$, we solve self-consistently 
Eqs.~(\ref{eq:kkk}) and (\ref{eq:ueq}) along with the equation for the  
free energy density, which has the following simplified expression
\begin{equation}
 f = \sum_i \left[ \sum_{k} n_i(k)
 \left( {k^2\over 2m_i} + {1\over 2}U_i(k) \right) - Ts_i \right] \:,
\label{fr_en}
\end{equation}
where
\begin{equation}
 s_i = - \sum_{k} \Big( n_i(k) \ln n_i(k) + [1-n_i(k)] \ln [1-n_i(k)] \Big)
\end{equation}
is the entropy density for component $i$ treated as a free gas with
spectrum $E_i(k)$. Finally, the pressure can be easily calculated as
the derivative of free energy with respect to the density, keeping fixed 
the temperature, i.e.,
\begin{equation}
 P = \rho^2 \Big( \frac{\partial (f/\rho)}{\partial \rho} \Big )_T \:. 
\end{equation}

\section{Composition and EOS of hot stellar matter}

For hot and neutrino trapped stellar matter containing only nucleons as 
relevant degrees of freedom, 
the composition is determined by the requirements of charge
neutrality and beta equilibrium, which read explicitly
\begin{eqnarray}
 && \sum_i q_ix_i + \sum_l q_l x_l = 0 \:,  \\
 && \mu_n - \mu_p = \mu_e - \mu_{\nu_e} = \mu_\mu + \mu_{\bar{\nu}_\mu} \:.
\label{beta:eps}
\end{eqnarray}
In the expression above, $x_i=\rho_i/\rho_B$ represents the baryon fraction for
the species $i$, $\rho_B$ the baryon density, and $q_i$ the electric charge.
The same holds true for the leptons, labelled by the subscript $l$.
It turns out that the dependence on the proton fraction can be approximated 
by the so-called parabolic approximation \cite{parabol}, which reads for the 
free energy as 
\begin{equation}
 \frac{F}{A}(\rho,x_p,T) \approx \frac{F}{A}(\rho,x_p=0.5,T) 
 + (1-2x_p)^2 {F_{\rm sym}}(\rho,T)
\end{equation}
being $F_{\rm sym}$ the symmetry energy
\begin{equation}
 {F_{\rm sym}}(\rho,T) \approx 
 \frac{F}{A}(\rho,x_p=0,T) - \frac{F}{A}(\rho,x_p=0.5,T) \:.
\end{equation}
In the parabolic approximation, 
the nucleon chemical potentials can be easily calculated as 
\begin{eqnarray}
 \mu_{n}(\rho,x_{p},T) &=& \frac{\partial f}{\partial \rho_n} =
 \left[ 1 + \rho \frac{\partial }{\partial\rho}
        - x_{p}\frac{\partial }{\partial x_{p}} \right] {f\over\rho} \:, 
\label{mun:eps}
\\ 
 \mu_{p}(\rho,x_{p},T) &=& \frac{\partial f}{\partial \rho_p} =
 \left[ 1 + \rho\frac{\partial}{\partial\rho}
          + (1-x_p)\frac{\partial}{\partial x_p} \right] {f\over\rho} \:,
\label{mup:eps}
\end{eqnarray}
whereas the chemical potentials of the noninteracting leptons 
are obtained by solving numerically the free Fermi gas model at finite 
temperature. 
Further details are given in Ref.~\cite{nic06}.

Because of trapping, the numbers of leptons per baryon
of each flavor $l=e,\mu$,
\begin{equation}
 Y_l = x_l - x_{\bar l} + x_{\nu_l} - x_{\bar{\nu}_l} \:,
\label{lepfrac:eps}
\end{equation}
are conserved on dynamical time scales.
Gravitational collapse
calculations of the white-dwarf core of massive stars indicate that at
the onset of trapping the electron lepton number is
$Y_e \approx 0.4$,
the precise value depending on the efficiency of electron capture
reactions during the initial collapse stage.
Moreover, since no muons are present when neutrinos become trapped,
the constraint
$Y_\mu = 0$
can be imposed.
We fix the $Y_l$ at those values in the calculations
for neutrino-trapped matter.

\subsection{Isentropic description of hot stellar matter}

\begin{figure}[t] %...........................................................
%\setcaptionmargin{5mm}
%\onelinecaptionstrue
\includegraphics[clip,scale=0.8]{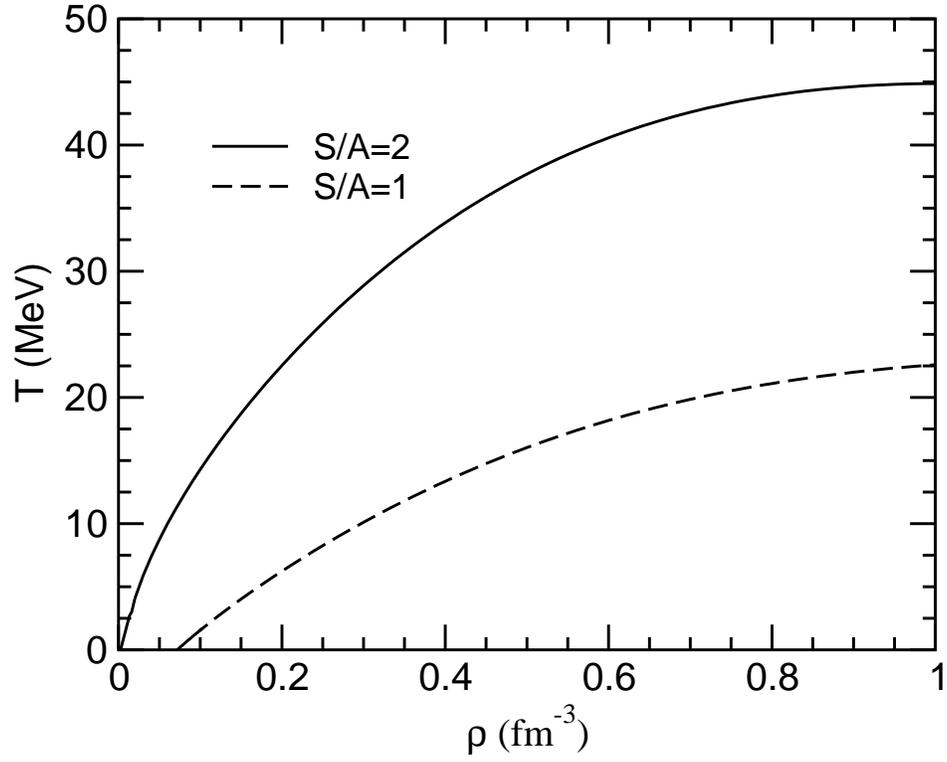}
%\captionstyle{normal}
\caption{
Temperature vs.~nucleon density for neutrino-trapped matter with
entropy S/A=1 (dashed line) and S/A=2 (solid line).} 
\label{f:Trho}
\end{figure}

As known from dynamical simulations, thermal effects result in an approximately
uniform entropy per baryon of 
$S/A \approx 1-2$ 
across the star \cite{burr}. 
Therefore, within the same theoretical framework, it may be useful to switch 
from an isothermal to an isentropic description of hot and trapped stellar 
matter, and compare the respective outcomes. In our approach, we proceed as 
follows. Once we determine the composition and the free energy per particle 
$F/A$ at a given density $\rho$,
for several values of the temperature $T$, we calculate the entropy per baryon 
(in units of the Boltzmann constant) 
from the thermodynamical relationship 
\begin{equation}
 \frac{S}{A} = -\Big( \frac{\partial F/A}{\partial T} \Big)_\rho \:.
\end{equation}
Hence, for fixed $S$, we find how 
the temperature changes as a function of the density $\rho$, as 
shown in Fig.~\ref{f:Trho}. 
The lower curve displays the calculation for $S/A=1$, 
whereas the upper curve is the calculation for $S/A=2$. 
We notice that the temperature is a monotonically increasing 
function of the density, and that the reachable values are larger 
for larger values of the entropy. We remark that the range of temperatures 
coincides with the values chosen for the isothermal description
of a protoneutron star \cite{nic06}. 

Following this procedure, we obtain the stellar composition and the 
corresponding free energy at a given nucleon density $\rho$.
Finally, we can calculate the total energy $E/A$ from the thermodynamical
relationship
\begin{equation}
 \frac{E}{A}(\rho,x_p,S) = \frac{F}{A}(\rho,x_p,S) 
 + T(\rho,x_p) \frac{S}{A} \:, 
\end{equation}
and the equation of state is
\begin{equation}
 P = \rho^2 \Big( \frac{\partial E/A}{\partial \rho} \Big )_S \:.
\end{equation}

The pressure vs.~density relationship is the fundamental input for
calculating stable configurations of a protoneutron star. These
can be obtained from the well-known hydrostatic equilibrium equations 
of Tolman, Oppenheimer, and Volkov
\cite{shapiro} 
for the pressure $P$ and the enclosed mass $m$
\bea
 {dP(r)\over dr} &=& -\frac{Gm(r)\eps(r)}{r^2}
 \frac{\big[ 1 + {P(r)\over\eps(r)} \big]
       \big[ 1 + {4\pi r^3P(r)\over m(r)} \big]}
 {1-{2Gm(r)\over r}} \:,
\label{tov1:eps}
\\
 \frac{dm(r)}{dr} &=& 4\pi r^{2}\eps(r) \:,
\label{tov2:eps}
\eea
once the EOS $P(\eps)$ is specified, being $\eps$
the total energy density ($G$ is the gravitational constant).
For a chosen central value of the energy density, the numerical integration of 
Eqs.~(\ref{tov1:eps}) and (\ref{tov2:eps}) provides the mass-radius relation.
For simplicity, we have attached a cold crust to the stellar core, 
by joining the hadronic equations of state above described with the 
ones by Negele and Vautherin \cite{nv} in the medium-density regime 
($0.001\;\mathrm{fm}^{-3}<\rho<0.08\;\mathrm{fm}^{-3}$), 
and the ones by Feynman, Metropolis, and Teller \cite{fey} 
and Baym, Pethick, and Sutherland \cite{baym}
for the outer crust ($\rho<0.001\;\mathrm{fm}^{-3}$). 

A more realistic treatment of the transition from the hot interior to the 
cold outer part has a dramatic influence on the mass-central density relation
in the region of low central density and low stellar masses, as discussed
in Ref.~\cite{Gondek}. 
However, the maximum mass region is not affected by the structure 
of this low-density transition region.

\section{Results and discussion}

Let us now discuss first the populations of beta-equilibrated
stellar matter, by solving the chemical equilibrium conditions given
by Eqs.~(\ref{beta:eps}), supplemented by electrical charge
neutrality, and baryon and lepton number conservation. 

\begin{figure}[t] %...........................................................
\includegraphics[clip,scale=0.8]{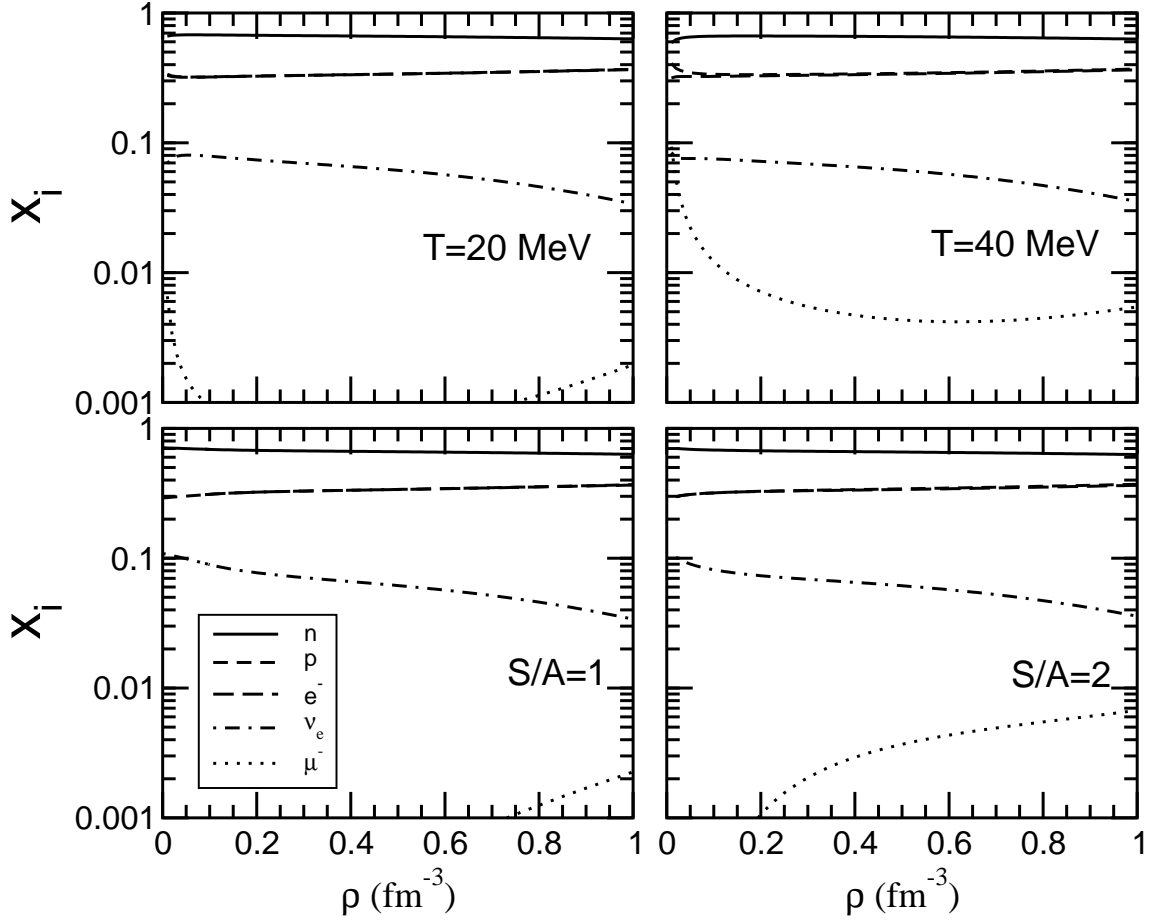}
\caption{
Relative populations for neutrino-trapped matter as a function of the 
nucleon density. Calculations are shown for the isothermal
case (upper panels), and for the isentropic one (lower panels).
} 
\label{f:xi}
\end{figure}

In Fig.~\ref{f:xi} we
show the particle fractions as a function of the nucleon density, for
different values of the temperature, respectively $T$=20,40 MeV
(upper panels), and entropy $S/A$=1,2 (lower panels). 
We do not observe sizeable differences between the two calculations, 
except for the neutrino fraction at extremely low densities, and
the muon percentage in the high temperature/entropy case. 

\begin{figure}[t] %...........................................................
\includegraphics[clip,scale=0.8]{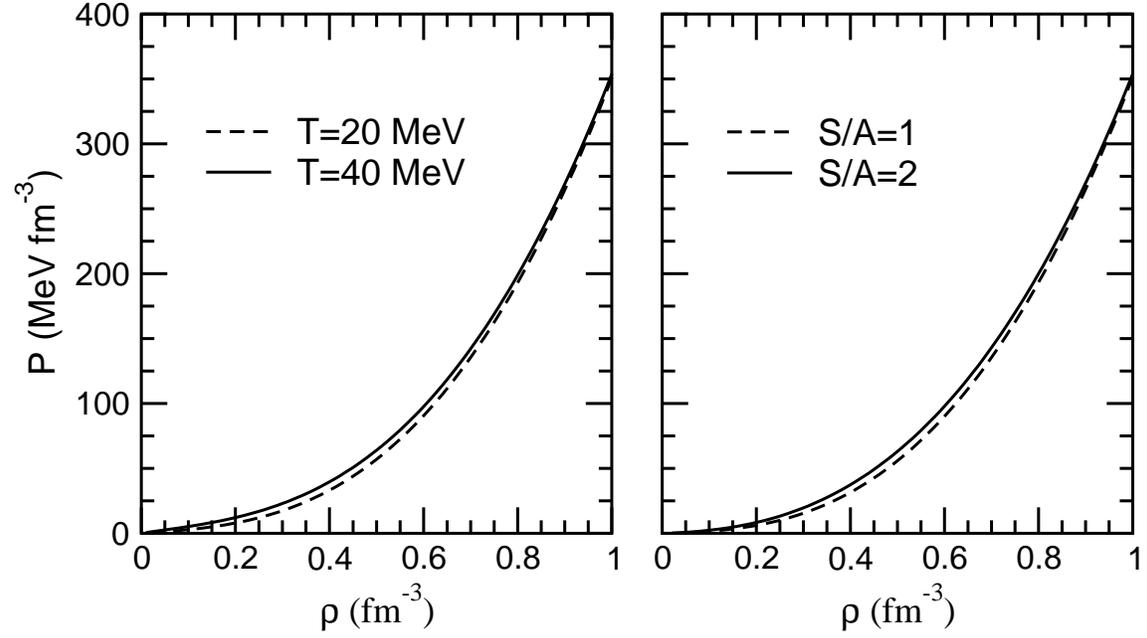} 
\caption{
Pressure as a function of the nucleon density for the isothermal case 
(left panel) and the isentropic one (right panel).
} 
\label{f:eos}
\end{figure}

The equation of state for beta-stable and neutrino trapped matter is displayed 
in Fig.~\ref{f:eos}, where the pressure is reported as a function of the
nucleon density. We observe that the equation of state stiffens with 
increasing $T$ or $S$ mostly in the medium-low density range.  
On the contrary, at large density thermal effects play a minor role, 
and curves of different $T$ or $S$ show a quite similar behavior, as
already found in Ref.~\cite{nic06}.

\begin{figure}[t] %...........................................................
\includegraphics[clip,scale=0.8]{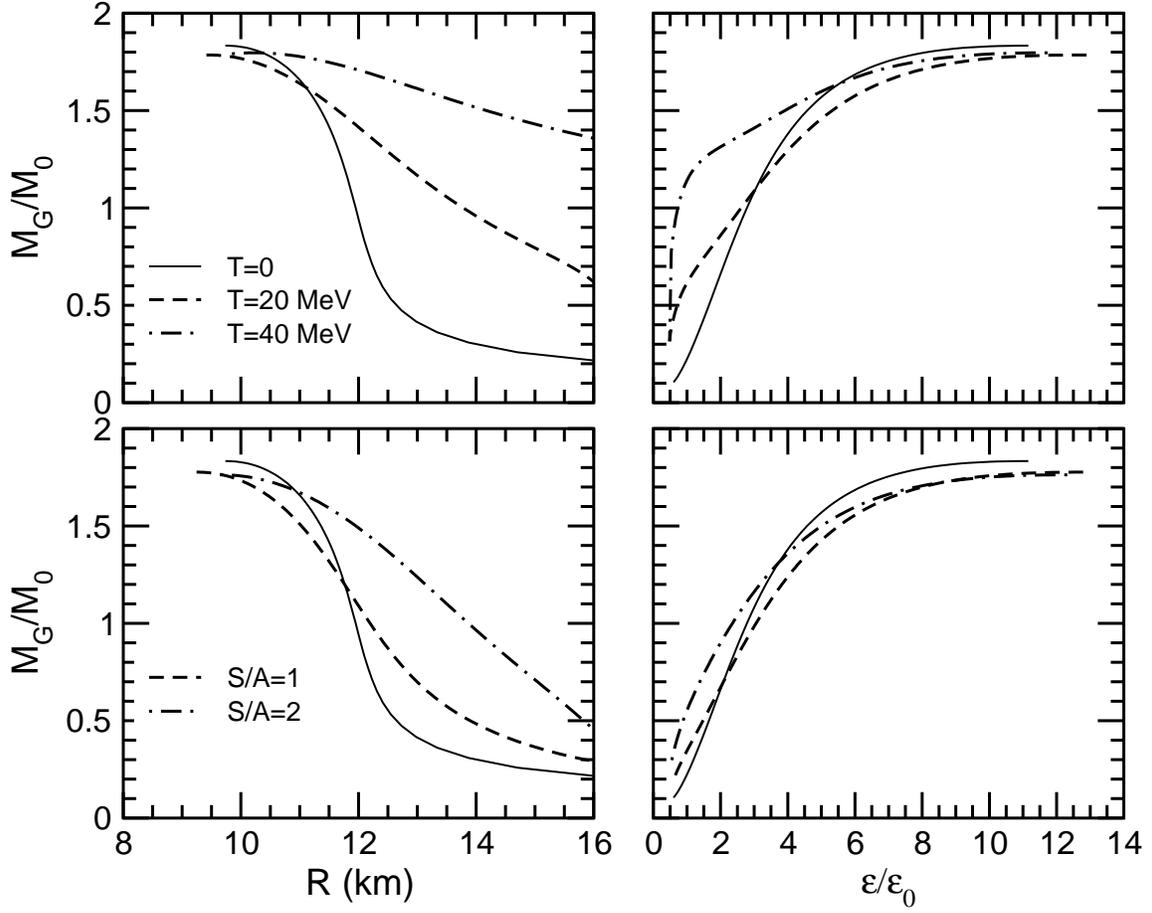} 
\caption{
Mass-radius (left panels) and mass-central density (right panels)
relations for isothermal (upper panels) and isentropic (lower panels)
conditions.
} 
\label{f:mr}
\end{figure}

Finally, in Fig.~\ref{f:mr} we display the gravitational mass as function of 
the radius (left panels), and the central energy density $\epsilon$ 
normalized with respect to the saturation value $\epsilon_0$ (right panels). 
The dashed (dot-dashed) line represents the calculations for 
the low (high) temperature or entropy case. 
We find that the values of the 
maximum mass are quite stable and almost independent on $S$ or $T$. 
In particular, we observe that the value of the maximum mass slightly 
decreases with increasing $S$.
The reason is that in hot and trapped stellar matter, the proton fraction 
increases with increasing $S$ or $T$ at large values of the nucleon density, 
and this leads to a softening of the equation of state, with consequent 
decrease of the maximum mass.
This result is in contrast with the findings of Prakash et al.~\cite{pra}, 
where the critical mass increases with increasing entropy.
This can be due to the fact that in Ref.~\cite{pra} the interaction is 
mostly local, with only a non-local correction, and therefore
only the kinetic part contains a dependence on temperature. 
In our calculations the whole interaction part is temperature-dependent, 
due to the intrinsic non-locality of the single-particle potentials, 
and therefore the temperature dependence can be quite different.

In the same figure we also show the results for cold neutrino-free 
neutron stars (thin solid line). 
One notes in particular that the maximum masses of the protostars 
are always slightly smaller than that of the neutron star,
which excludes a delayed collapse while cooling down \cite{pra,pons}.

\section{Conclusions}

We have modeled nucleonic protoneutron stars using a realistic BBG
equation of state and assuming either isothermal or isentropic conditions.
Heavy protostars are fairly insensitive to that choice 
and their maximum masses are slightly smaller than that of the cold 
neutron star.

%%%%%%%%%%%%%%%%%%%%%%%%%%%%%%%%%%%%%%%%%%%%%%%%%%%%%%%%%%%%%%%%%%%%%%%%%%%%%

%%%%%%%%%%%%%%%%%%%%%%%%%%%%%%%%%%%%%%%%%%%%%%%%%%%%%%%%%%%%%%%%%%%%%%%%%%%%%
\newpage

 %................................................................

 %................................................................

 %................................................................

 %................................................................

\end{document}